\documentclass[prc,10pt,twocolumn,showkeys,
nofootinbib,
 amsmath,amssymb,
 aps,
]{revtex4-1}
\usepackage{yfonts}
\usepackage{graphicx}
\usepackage{dcolumn}
\usepackage{bm}
\usepackage{nicefrac}
\usepackage{slashed}
\usepackage{xcolor}
\definecolor{darkblue}{RGB}{0,0,196}
\definecolor{darkgreen}{RGB}{0,120,0}
\definecolor{magenta}{RGB}{255,0,255}
\usepackage[colorlinks=true,linktocpage=true,linkcolor=darkblue,citecolor=red,urlcolor=darkblue]{hyperref}
\usepackage{hyperref}
\usepackage{cancel}
\newcommand{\beq}{\begin{equation}}
\newcommand{\eeq}{\end{equation}}

\newcommand{\bea}{\begin{eqnarray}}
\newcommand{\eea}{\end{eqnarray}}

\newcommand{\bel}[1]{\begin{eqnarray}\label{#1}}
\newcommand{\eel}{\end{eqnarray}}

\def\LB{\left(}
\def\RB{\right)}

\newcommand{\EQ}[1]{Eq.~(\ref{#1})}
\newcommand{\EQn}[1]{(\ref{#1})}

\newcommand{\EQSTWO}[2]{Eqs.~(\ref{#1})~and~(\ref{#2})}

\newcommand{\CIT}[1]{Ref.~\citep{#1}} 
\newcommand{\CITn}[1]{\citep{#1}} 

\newcommand{\dd}{\mathrm{d}}

\def\epsLmnab{\epsilon_{\mu\nu\alpha\beta}}

\newcommand{\Av}{{\boldsymbol A}} 
\newcommand{\Bv}{{\boldsymbol B}} 
\newcommand{\Cv}{{\boldsymbol C}}

\newcommand{\av}{{\boldsymbol a}} 
\newcommand{\bv}{{\boldsymbol b}} 
\newcommand{\bvp}{{\boldsymbol b}^\prime} 
\newcommand{\ev}{{\boldsymbol e}}
\newcommand{\evp}{{\boldsymbol e}^\prime}

\newcommand{\vv}{{\boldsymbol v}}
\newcommand{\pv}{{\boldsymbol p}}
\newcommand{\pvp}{{\boldsymbol p}^\prime}
\newcommand{\php}{{\hat {\boldsymbol p}}^\prime}

\newcommand{\sv}{{\boldsymbol s}}
\newcommand{\xv}{{\boldsymbol x}}

\newcommand{\trf}{{\rm tr_4}}

\newcommand{\f}[2]{\frac{#1}{#2}}
\newcommand{\onehalf}{{\nicefrac{1}{2}}} 
 
\newcommand{\threefourths}{{\nicefrac{3}{4}}} 

\def\spin{\,\textgoth{s:}}
\def\spinl{|{\boldsymbol s}_*|}

\def\omnL{\omega_{\mu\nu}}
\def\omnU{\omega^{\mu\nu}}

\def\omnLD{{\tilde \omega}_{\mu\nu}}
\def\omnUD{\tilde {\omega}^{\mu\nu}}

\begin{document}

\title{Application range of perfect spin hydrodynamics}

\author{Zbigniew Drogosz}
\email{zbigniew.drogosz@alumni.uj.edu.pl}
\affiliation{Institute of Theoretical Physics, Jagiellonian University, PL-30-348 Krak\'ow, Poland}

\author{Wojciech Florkowski}
\email{wojciech.florkowski@uj.edu.pl}
\affiliation{Institute of Theoretical Physics, Jagiellonian University, PL-30-348 Krak\'ow, Poland}

\author{Valeriya Mykhaylova}
\email{valeriya.mykhaylova@uj.edu.pl}
\affiliation{Institute of Theoretical Physics, Jagiellonian University, PL-30-348 Krak\'ow, Poland}

\date{\today}

\begin{abstract}
The application range of perfect spin hydrodynamics is studied in two cases: one based on the classical spin description and the other using a quantum spin density matrix (Wigner function). Different forms of the conditions connecting the components of the spin polarization tensor, particle mass, temperature, and hydrodynamic flow are introduced, and their mutual relations are explained. The results obtained are important for practical applications of spin hydrodynamics to model heavy-ion collisions. 
\end{abstract}

                              
\maketitle

\section{Introduction}

Spin-polarization measurements of various hadrons produced in relativistic heavy-ion collisions (for example, of the $\Lambda$ hyperons and vector mesons) have recently attracted much interest~\CITn{Liang:2004ph, Liang:2004xn, STAR:2017ckg, STAR:2018gyt, STAR:2019erd, ALICE:2019aid}. For a current experimental review of this broad topic, we refer the reader to~\CITn{Niida:2024ntm}. The experimental data naturally inspired a wide range of theoretical studies~\CITn{Becattini:2009wh, Becattini:2011zz, Becattini:2013fla, Montenegro:2017rbu, Florkowski:2017ruc, Florkowski:2017dyn, Florkowski:2018ahw, Hattori:2019lfp, Weickgenannt:2019dks, Florkowski:2019qdp, Bhadury:2020puc, Montenegro:2020paq, Weickgenannt:2020aaf, Shi:2020htn, Bhadury:2020cop, Fukushima:2020ucl, Li:2020eon, Singh:2020rht, Gallegos:2021bzp, Weickgenannt:2021cuo, She:2021lhe, Hongo:2021ona, Hu:2021pwh, Singh:2022ltu, Weickgenannt:2022zxs, Wagner:2022amr, Dey:2023hft, Weickgenannt:2023nge, Kumar:2023ojl, Wagner:2024fhf, Wagner:2024fry, Dey:2024cwo, She:2024rnx, Huang:2024ffg, Bhadury:2025fil, Dey:2025wqw, Weickgenannt:2024ibf}, which in particular have led to the concept of spin hydrodynamics~\CITn{Florkowski:2017ruc}. The latter framework implements spin degrees of freedom in the formalism of ``ordinary'' relativistic hydrodynamics that is known as one of the most effective tools used to interpret the heavy-ion data~\CITn{Ollitrault:2007du, Romatschke:2009im, Florkowski:2010zz, Gale:2013da, Jaiswal:2016hex}. Although the formalism of spin hydrodynamics has been developed for several years now, fully 3D numerical codes where the spin degrees of freedom follow their own dynamic equations have been formulated only very recently~\CITn{Singh:2024cub, Sapna:2025yss}. Moreover, the construction of the spin hydrodynamics formalism itself is not yet complete, as several formal issues still require clarification. One of such problems is the applicability range of perfect spin hydrodynamics, which is thoroughly analyzed in this paper. 

The main object of interest in spin hydrodynamics is the spin polarization tensor $\omega_{\alpha\beta}$ that can be interpreted as the dimensionless ratio of the spin chemical potential $\Omega_{\alpha\beta}$ to the temperature~$T$, $\omega_{\alpha\beta} = \Omega_{\alpha\beta}/T$. Formal studies of spin hydrodynamics and its practical applications are mostly restricted to the case where the components of $\omega_{\alpha\beta}$ are small (we note that the expansion in $\omega_{\alpha\beta}$ is meaningful since the components $\omega_{\alpha\beta}$ are dimensionless). The main exceptions are: the review article~\CITn{Florkowski:2018fap}, where the classical description of spin was formulated, and the three more recent publications~\CITn{Florkowski:2024bfw, Drogosz:2024gzv, Bhadury:2025boe}, where the generalized thermodynamic identities were introduced and a novel form of the equilibrium Wigner function was proposed. 

In~\CIT{Bhadury:2025boe}, a criterion was given for the applicability of perfect spin hydrodynamics (describing a relativistic gas of spin-$\onehalf$ particles), which imposes an upper bound for the magnitude of the components of the spin polarization tensor,
\begin{eqnarray}
\f{1}{2} \, \omega_{\rm max} \, (2 + \sqrt{2}) 
\sqrt{\f{1+v}{1-v}}< \f{m}{T}.  
\label{eq:firstcond}
\end{eqnarray}
Here $\omega_{\rm max}$ is the largest absolute value of the components of the spin polarization tensor $\omega_{\alpha\beta}$ defined in the laboratory (hereafter LAB) frame, $m$ is the mass of the particles and $v$ denotes the magnitude of the flow three-vector (defined again in the LAB). It was also shown in~\CITn{Bhadury:2025boe} that~\EQ{eq:firstcond} is naturally fulfilled by the physical conditions encountered in heavy-ion collisions. 

In this work, we elucidate the origin of the condition~\EQn{eq:firstcond} and derive finer constraints that involve all components of the spin polarization tensor. Our description refers to the classical spin approach~\CITn{Florkowski:2018fap} and the quantum framework based on the Wigner function~\CITn{Bhadury:2025boe}.   We also discuss in parallel the conditions formulated in the fluid rest frame (LRF) and in an arbitrary frame that we identify with LAB. The results presented herein will be very useful in future applications of spin hydrodynamics to model heavy-ion collisions. 

\smallskip
Before we discuss our methods and results, it is important to emphasize two points. First, by the applicability range of perfect spin hydrodynamics, we mean the internal consistency of this framework. We do not address here the issue of including dissipation. Second, by a perfect spin fluid we understand a system in local equilibrium defined by the conservations of the baryon number, energy, linear momentum, spin part of the total angular momentum, and entropy. The effects of the spin-orbit interaction leading to a transfer between the spin and orbital parts of the total angular momentum can be added by the dissipative terms, as explained in~\CITn{Florkowski:2024bfw, Drogosz:2024gzv}. They are not related to the problems discussed in this work.

\smallskip
{\it Notation and conventions --} Throughout the text, we use natural units $\hbar = c = k_{\rm B} = 1$. For the Levi-Civita symbol $\epsilon^{\mu\nu\alpha\beta}$, we follow the convention \mbox{$\epsilon^{0123} =-\epsilon_{0123} = +1$}. We use the mostly negative metric convention, \mbox{$g_{\mu\nu} = \textrm{diag}(+1,-1,-1,-1)$}. The trace over the spinor indices is denoted by $\trf$. The scalar product of two four-vectors $a$ and $b$ reads $a \cdot b = a^0 b^0 - \av \cdot \bv$, where the three-vectors are indicated in bold. The tilde diacritic is used to denote dual tensors, which are obtained from rank two antisymmetric tensors $a_{\mu\nu}$ by contraction with the Levi-Civita symbol and division by a factor of two. For example, ${\tilde a}_{\mu\nu} = (1/2) \,\epsLmnab \,  a^{\alpha \beta}$. The inverse transformation is $a^{\rho \sigma} = -(1/2) \, \epsilon^{\rho \sigma \mu \nu}
{\tilde a}_{\mu \nu}$.

\medskip
\section{Classical spin description} 

\subsection{Extended phase space}

In the classical-spin approach we use distribution functions $f(x,p,s)$ in the extended phase space that includes the spin four-vector $s^\mu$ along with the spacetime coordinates $x^\mu = (t, \xv)$ and the four-momentum $p^\mu = (E_p, \pv)$. Throughout this work, we assume that particles are on the mass shell, hence $E_p =\sqrt{m^2 + \pv^2}$.

The spin four-vector $s^\mu$ is orthogonal to the particle four-momentum, $p \cdot s = 0$, and satisfies the normalization condition $s^2 = -\spin^2$, where $\spin^2 = \threefourths$ is the eigenvalue of the Casimir operator for the SU(2) group. Following the works by Mathisson~\CITn{Mathisson:1937zz,2010GReGr..42.1011M}, we define the internal angular momentum of a particle by the equation $ s^{\alpha \beta} = (1/m) \epsilon^{\alpha\beta\gamma\delta} p_\gamma s_\delta$. In the particle rest frame~(PRF), where $p^\mu = (m,0,0,0)$, the four-vector $s^\alpha$ has only space components, $s^\alpha = (0,\sv_*)$, with the normalization~$\spinl = \spin$.

The allowed spin configurations can be included by using a covariant measure~\CITn{Florkowski:2018fap}
\bea
\int \dd S \ldots = \f{m}{\pi \spin}  \, \int \dd^4s \, \delta(s \cdot s + \spin^2) \, \delta(p \cdot s) \ldots \,.
\label{eq:measureS}
\eea
The two delta functions in \EQ{eq:measureS} take care of the normalization and orthogonality conditions, while the prefactor $m/(\pi \spin)$ is chosen to obtain the normalization that accounts for the spin degeneracy factor of spin-$\onehalf$ particles,
\bea
\int \dd S = \f{m}{\pi \spin}  \int \, \dd^4s \, \delta(s \cdot s + \spin^2) \, \delta(p \cdot s) = 2.
\label{eq:intS}
\eea
The measure in the momentum space, denoted by $\dd P$, is defined as
\bel{eq:measureP}
\int \dd P \ldots = \int \frac{\dd^3p}{(2 \pi )^3 E_p} \ldots \,\,.
\eel
We note that both $\dd P$ and $\dd S$ are Lorentz invariant.

\subsection{Local equilibrium function}

The local equilibrium distribution functions for particles $(+)$ and antiparticles $(-)$ in the extended phase space have the form 
\bel{eq:classEQ}
f^\pm(x,p,s) = \exp\left(\pm \xi - p \cdot \beta + \frac{1}{2} \omega_{\alpha \beta} s^{\alpha \beta} \right),
\eel
where $\beta^\mu= u^\mu/T$, $u^\mu = \gamma (1, \vv )$ is the fluid four-velocity, and $\xi = \mu/T$, where $\mu$ is the baryon chemical potential. Equation~\EQn{eq:classEQ} is valid for the Boltzmann statistics. In~this case, the fugacity factor $\exp(\pm \xi)$ is irrelevant to our analysis; therefore, we set \mbox{$\xi = \mu/T =0$} and consider only particles in the following discussion.

The spin-polarization tensor $\omega_{\alpha\beta}$ appearing in~\EQn{eq:classEQ} can be expressed in terms of the electric- and magneticlike three-vectors, \mbox{$\ev = (e^1,e^2,e^3)$} and $\bv = (b^1,b^2,b^3)$, as
\bel{omeb}
\omega_{\mu\nu} = 
\begin{bmatrix}
0     &  e^1 & e^2 & e^3 \\
-e^1  &  0    & -b^3 & b^2 \\
-e^2  &  b^3 & 0 & -b^1 \\
-e^3  & -b^2 & b^1 & 0
\end{bmatrix},
\eel
where the sign conventions follow those used for the Faraday tensor in~\cite{Jackson:1998nia}.

\subsection{Scalar density – classical version}

An object of central importance for our analysis is the scalar density
\begin{align}\begin{split}
\label{eq:SDC}
n_{\rm cl} &= \int \dd P \int \dd S  \exp\left(- p \cdot \beta + \frac{1}{2} \omega_{\alpha \beta} s^{\alpha \beta} \right) \\
&= \int \dd P   \exp\left(- p \cdot \beta \right)
\int \dd S \exp\left(\frac{1}{2} \omega_{\alpha \beta} s^{\alpha \beta} \right).
\end{split}\end{align}
All other tensors of interest can be obtained by taking derivatives of $n_{\rm cl}$ with respect to $\beta^\mu$ and/or $\omega_{\alpha\beta}$. The integral over the spin space is analytic and equals~\CITn{Florkowski:2018fap}
\bel{eq:sinh}
\int \dd S \, \exp\LB  \f{1}{2}  \omega_{\alpha \beta} s^{\alpha\beta}\RB  
&=& \f{2 \sinh( \spin |\bv_*|)}{ \spin \, |\bv_*|},
\eel
where $\bv_*$ is the magneticlike part of the spin polarization tensor boosted to the particle rest frame
\bel{eq:bstar}
\bv_* &=& \f{1}{m} \left(  E_p \, \bv - \pv \times \ev - \f{\pv \cdot \bv}{E_p + m} \pv \right).
\eel

Since $n_{\rm cl}$ is a Lorentz scalar, it can be calculated in the local rest frame of the fluid element (LRF), where \mbox{$u'^{\mu} = (1,0,0,0)$}. Denoting quantities in LRF by primes, we obtain
\begin{align}\begin{split}
\label{eq:SDC1}
\!\!\!\! n_{\rm cl}
&= \int \dd P^\prime   \exp\left(- \f{E_{p^\prime}}{T} \right) \int \dd S^\prime \exp\left(\frac{1}{2} \omega_{\alpha \beta}^\prime s^{\prime \,  \alpha \beta} \right)
 \\
&=  2 \int \dd P^\prime   \exp\left(- \f{E_{p^\prime}}{T} \right)
\f{\sinh( \spin |\bv^\prime_*|)}{ \spin \, |\bv^\prime_*|},
\end{split}\end{align}
where we used~\EQ{eq:sinh} with
\bel{eq:bstarprime}
\bv^\prime_* &=& \f{1}{m} \left[  E_{p^\prime} \, \bv^\prime - \pv^\prime \times \ev^\prime - \f{\pv^\prime \cdot \bv^\prime}{E_{p^\prime} + m} \pv^\prime \right].
\eel
Here $\bv^\prime$ and $\ev^\prime$ are the $\bv$ and $\ev$ fields transformed to~LRF~\cite{Jackson:1998nia},
\begin{align}
\label{eq:fieldsJacksone}
\ev^\prime &= \gamma \left( \ev + \vv \times \bv \right) - \frac{\gamma^2}{\gamma+1} \vv (\vv \cdot \ev),\\
\bv^\prime &= \gamma \left( \bv - \vv \times \ev \right) - \frac{\gamma^2}{\gamma+1} \vv (\vv \cdot \bv).\label{eq:fieldsJacksonb}
\end{align}

\subsection{Constraining the spin polarization components}

From~\EQ{eq:SDC1} it is easy to conclude that the existence of the integral defining $n_{\rm cl}$ is controlled by the behavior of the difference
\begin{eqnarray}
\label{eq:tau}
\tau \equiv -\f{E_{p^\prime}}{T} + \spin |\bv^\prime_*|
\end{eqnarray}
at large values of the three-momentum $|\pv^\prime|$. For large~$|\pvp|$, \EQSTWO{eq:bstarprime}{eq:tau} yield
\begin{equation}
\tau = -\f{|\pvp|}{T} + \spin \f{|\pvp|}{m} \left| \bvp - \php \times \evp - (\php \cdot \bv)\php \right|,
\end{equation}
with hats denoting unit vectors. The integral (\ref{eq:SDC1}) converges if $\tau < 0$ for every choice of $\php$. Therefore, to find whether it converges for given $m$, $T$, $\bvp$, and $\evp$, we have to maximize the quantity
\begin{equation}
\mathcal{Q}' \equiv \left| \bvp - \php \times \evp - (\php \cdot \bvp)\php \right|
\end{equation}
with respect to $\php$. Squaring both sides and making use of elementary trigonometric identities as well as of the cyclic property of the mixed product $\Av \cdot (\Bv \times \Cv)$ leads to
\begin{equation}\label{eq:q2}
\mathcal{Q}'^2 = {\bvp}^2 \sin^2 \theta_{bp}+ {\evp}^2 \sin^2 \theta_{ep} - 2 \php \cdot (\evp \times \bvp),
\end{equation}
where $\theta_{bp}$ and $\theta_{ep}$ denote the angles between $\php$ and the vectors $\bvp$ and $\evp$, respectively. Now, it can be readily seen that to maximize this expression, $\php$ should be chosen as the unit vector antiparallel to $\evp \times \bvp$.\footnote{If $\evp \times \bvp = 0$, i.e. the vectors are parallel or one of them is zero, then $\php$ can be chosen as any vector perpendicular to (nonzero) $\bvp$ or $\evp$.} Then, the sines are equal to $\pm1$, and
\begin{equation}
\mathcal{Q}'^2 = {\bvp}^2 + {\evp}^2 + 2 |\evp \times \bvp|.
\end{equation}
Hence, the criterion for convergence of (\ref{eq:SDC1}), $\max(\tau) < 0$, takes the form
\bel{eq:criterion}
\spin \sqrt{ {\bvp}^2 + {\evp}^2 + 2 |\evp \times \bvp|} < \f{m}{T}.
\eel
Equation~\EQn{eq:criterion} is our first important result. We note that it differs from the criterion presented in \cite{Rajeev:2025QM} by the appearance of the cross product and by its ``primed'' form, meaning that it can be applied in any frame if the primed (LRF) components are expressed by those in LAB.

\subsection{Maximum-magnitude criteria}\label{subs:maxmag}
If the components $e^i \,\, (i\!=\!1,2,3)$ are restricted to lie in the interval $-\omega_{\rm max} \leq e^i \leq \omega_{\rm max}$ for a certain positive $\omega_{\rm max}$, and the same holds for the components \mbox{$b^i \,\, (i\!=\!1,2,3)$}, then one can prove (see Appendix~\ref{sec:proofQ}) that $\sqrt{ {\bv}^2 + {\ev}^2 + 2 |\ev \times \bv|}$ is bounded from above by \mbox{$\omega_{\rm max} (2+ \sqrt 2)$}. Therefore, based on (\ref{eq:criterion}), we can formulate a maximum-magnitude criterion for $\omega_{\rm max}^\prime$, namely
\begin{equation}\label{eq:maxmag}
\spin \ \omega_{\rm max}^\prime (2 + \sqrt 2) < 
\f{m}{T}.
\end{equation}
We note that~\EQn{eq:maxmag} is more restrictive than~\EQn{eq:criterion}, i.e.~\EQn{eq:maxmag}~implies \EQn{eq:criterion}. However, it may be simpler to use, especially as it can easily be transformed into a condition that is valid in LAB. 

The upper bound on $\omega_{\rm max}^\prime$ given $\omega_{\rm max}$ and $v$ can be found as follows. Let us assume that $|\ev| \geq |\bv|$. We rotate the coordinate frame so that $\ev = (\omega_{\rm max},0,0)$ and seek the choice of $\bv$ and $\vv$ that maximizes ${e^\prime}^1 = \omega_{\rm max}^\prime$. We~find that the projection of the last term in \EQn{eq:fieldsJacksone} on $\ev$ is nonpositive regardless of the choice of $\vv$,
\begin{equation}
- \frac{\gamma^2}{\gamma+1} (\vv \cdot \ev)^2 \leq 0.
\end{equation}
Therefore, in the extremal case, this term should be zero, whereas the projection of the second term, $\gamma \vv \times \bv$, on $\ev$ should be as large as possible. It is easy to verify that the choice that maximizes 
${e^\prime}^1$ is $\bv \perp \ev$, $\vv \perp \ev$, $\vv \perp \bv$, $|\bv| = |\ev|$, with $\vv \times \bv$ parallel to $\ev$. Then,
\begin{equation}
\ev^\prime = \gamma (1+v)\ev = \sqrt{\f{1+v}{1-v}}\ev,
\end{equation}
and the criterion \EQn{eq:maxmag} takes the form\footnote{The result for the case $|\bv| \geq |\ev|$ is the same, and the only required modification of the argument is to use \EQ{eq:fieldsJacksonb} and choose $\vv$ and $\ev$ such that $\vv \times \ev$ is antiparallel to $\bv$.}
\begin{equation}\label{eq:maxmag2}
\spin \, \omega_{\rm max} \left(2 + \sqrt{2} \right)\sqrt{\f{1+v}{1-v}} < \f{m}{T}.
\end{equation}
This criterion is coarser (more restrictive) than that derived in the previous subsection, as it corresponds to the worst-case arrangement of the vectors $\ev$, $\bv$, and $\vv$, but it may be easier to apply in practical calculations, as it uses only two variables, $\omega_{\rm max}$ and $v$, instead of the nine components of the considered vectors.

\section{Wigner function approach}

Equation \EQn{eq:maxmag2} has exactly the same form as~\EQ{eq:firstcond} provided that the factor $\spin$ is replaced by $1/2$. Since~\EQ{eq:firstcond} was derived within the framework based on the Wigner function, we first recall the main elements of this approach.

\subsection{Local equilibrium}

A very recent paper \cite{Bhadury:2025boe} introduced a local equilibrium Wigner function defined by the expression
\bel{XpmNEW3}
X(x,p) =  \exp\left[- \beta_\mu(x) p^\mu + \gamma_5 \slashed{a} \right],
\eel
where
\bel{eq:amu}
a_\mu(x,p) = -\frac{1}{2 m} {\tilde \omega}_{\mu\nu}(x)p^\nu.
\eel
The tensor ${\tilde \omega}_{\mu\nu}$ is the dual polarization tensor
\bel{omdeb}
\omnLD = 
\begin{bmatrix}
	0       &  b^1 & b^2 & b^3 \\
	-b^1  &  0    & e^3 & -e^2 \\
	-b^2  &  -e^3 & 0 & e^1 \\
	-b^3  & e^2 & -e^1 & 0
\end{bmatrix}.
\label{eq:omegaebd}
\eel
We note that the following identities hold
\begin{align}\begin{split} \label{om3}
\f{1}{2} \omnL \omnU &= -\f{1}{2} \omnLD \omnUD 
=   \bv \cdot \bv - \ev \cdot \ev, \\
\omnLD \omnU &= -4 \ev \cdot \bv . 
\end{split}\end{align}
The orthogonality condition $a_\mu p^\mu$ may be treated as a version of the Frenkel condition~\CITn{Frenkel:1926zz} that is frequently used in spin hydrodynamics in different formulations.

\subsection{Scalar density – quantum version }

The results for conserved tensors (including baryon current, energy-momentum tensor, spin tensor, and entropy current) obtained with classical and quantum descriptions have been shown to agree up to the second order in $\omega$. Interestingly, in this case, the integrals over allowed classical spin configurations are equivalent to the calculation of the traces over the spinor space. 
 
The scalar density calculated with the quantum treatment of spin is given by the formula
\bel{eq:SDQ}
n_{\rm qt} = \f{1}{2} \, \trf  \int \! \dd P  \exp\left(- p \cdot \beta + \gamma_5 \slashed{a} \right),
\eel
where the prefactor $1/2$ guarantees that the quantum formula agrees with the classical one for the spinless case. To proceed further, it is convenient to make use of the identity
\begin{equation}\label{eq:expa}
\exp\!\left( \gamma_5 \slashed{a} \right)\!\!=\!\cosh\!\sqrt{-a^2}
\!\left[\!1+\!\frac{\gamma_5 \slashed{a}}{\sqrt{-a^2}} \tanh\!\sqrt{-a^2} \right]
\end{equation}
that yields
\begin{equation}\label{eq:SDQ1}
n_{\rm qt} = 2  \int \! \dd P  \exp\left(- p \cdot \beta \right) \,\cosh\!\sqrt{-a^2}.
\end{equation}
%

\subsection{Constraining the spin polarization components}

Since the integral \EQn{eq:SDQ1} is a scalar, we can write it down in the LRF,
\begin{eqnarray}
n_{\rm qt} &=& 2  \int \! \dd P^\prime  \exp\left(- p' \cdot \beta \right) \,\cosh\!\sqrt{-a'^2}.
\end{eqnarray}
This integral converges if for large $|\pvp|$ the expression
\begin{equation}
-\f{p'}{T} + \sqrt{-a'^2}
\end{equation}
is negative for every direction of $\pvp$.
This looks similar to \EQn{eq:tau} and, in fact, it can be easily verified that
\begin{align}\begin{split}
-a^{\prime \,2} &= \f{|\pv'|}{4m^2} \left[{\bvp}^2 - ({\hat \pv}^\prime \cdot {\bv}^\prime)^2 + ({\hat \pv}^\prime \times \evp)^2 \!-\! 2 {\hat \pv}^\prime \cdot (\evp \times \bvp) \right]\\
&=\f{|\pv'|}{4m^2} \left[{\bvp}^2 \sin^2 \theta_{bp}+ {\evp}^2 \sin^2 \theta_{ep} - 2 \,{\hat \pv}^\prime \cdot (\evp \times \bvp)\right],
\end{split}\end{align}
with $\theta_{bp}$ and $\theta_{ep}$ defined analogously as in \EQ{eq:q2}. Then, the same argument as below \EQ{eq:q2} leads to the convergence criterion
\bel{eq:criterionW}
\f{1}{2} \sqrt{ {\bvp}^2 + {\evp}^2 + 2 |\evp \times \bvp|} < \f{m}{T}.
\eel

\subsection{Maximum-magnitude criteria}

Since the inequalities (\ref{eq:criterion}) and (\ref{eq:criterionW}) differ only by a multiplicative constant, the entirety of Sec.~\ref{subs:maxmag} is valid here as well. In the Wigner case, the maximum-magnitude criteria analogous to (\ref{eq:maxmag}) and (\ref{eq:maxmag2}) take the form
\begin{gather}\label{eq:maxmagW}
\f{1}{2} \omega_{\rm max}^\prime (2 + \sqrt 2) < 
\f{m}{T},\\
\f{1}{2} \omega_{\rm max} \left(2 + \sqrt{2} \right)\sqrt{\f{1+v}{1-v}} < \f{m}{T}.
\end{gather}

\begin{table*}[t]
\begin{center}
\caption{Applicability criteria for different sets of available variables and for two different formulations of perfect spin hydrodynamics (the second column -- the approach based on the classical spin description, the third column -- the approach using the Wigner function; $\omega_{\rm max} = \max(|e^i|,|b^i|), \ i=1,2,3$). }\label{tab:v}\vspace{4pt}
\begin{tabular}{|ccc|}
\hline
\multicolumn{1}{|c|}{$\quad$ Variables $\quad$} & \multicolumn{1}{c|}{Classical} & Wigner \\ [1.5ex]\hline 
\multicolumn{1}{|c|}{$\omega_{\rm max}, v$} & \multicolumn{1}{c|}{$\displaystyle \spin \, \omega_{\rm max}(2 + \sqrt 2)\sqrt{\f{1+v}{1-v}}   < \f{m}{T}$} & $ \displaystyle \f{1}{2}\omega_{\rm max}(2 + \sqrt 2)\sqrt{\f{1+v}{1-v}}   < \f{m}{T}$ \\[10pt] \hline
\multicolumn{1}{|c|}{$\ev, \bv, v$} & \multicolumn{1}{c|}{$ \displaystyle \quad \spin \, \sqrt{ {\bv}^2 + {\ev}^2 + 2 |\ev \times \bv|} \, \sqrt{\f{1+v}{1-v}} < \f{m}{T} \quad$} & $\displaystyle \quad \f{1}{2} \sqrt{ {\bv}^2 + {\ev}^2 + 2 |\ev \times \bv|} \, \sqrt{\f{1+v}{1-v}} < \f{m}{T} \quad$ \\[10pt] \hline
\multicolumn{1}{|c|}{$\ev, \bv, \vv$} & \multicolumn{1}{c|}{$ \displaystyle \spin \, \sqrt{ {\bvp}^2 + {\evp}^2 + 2 |\evp \times \bvp|} < \f{m}{T}$} & $ \displaystyle \f{1}{2} \sqrt{ {\bvp}^2 + {\evp}^2 + 2 |\evp \times \bvp|} < \f{m}{T}$ \\[10pt] \hline
\end{tabular}
\end{center}
\end{table*}

\section{Summary and conclusions}

In this work, we have constructed several criteria that constrain the application of perfect spin hydrodynamics in the formulations defined in~\CIT{Florkowski:2018fap, Florkowski:2024bfw, Drogosz:2024gzv, Bhadury:2025boe}. These criteria are a general restriction on the magnitude of the allowed values of the spin-polarization tensor components. The maximum values allowed depend on the mass-to-temperature ratio, $m/T$, and the value of the hydrodynamic flow $v$. We have verified all the conditions by performing numerical calculations/integrations.

The specific form of the criteria depends on the set of variables given. A full collection of our results is provided in Table~\ref{tab:v}. If we have complete data on the vectors $\ev$, $\bv$ and $\vv$ (in the laboratory frame), we can apply the most exact version (third row in Table~\ref{tab:v}), which uses the Lorentz-transformed fields; see \EQSTWO{eq:fieldsJacksone}{eq:fieldsJacksonb}. If only the~magnitude of $\vv$ is given but not its direction relative to $\bv$ and $\ev$, we can notice that the left-hand side is bounded from above by an expression that uses non-transformed fields multiplied by a Doppler effect-like factor (second row in Table~\ref{tab:v}). Finally, given only $v$ and $\omega_{\rm max}$ (the largest component of the vectors $\ev$ and $\bv$), we can use the criterion that involves only these two variables and is the upper bound of the two finer criteria (first row in Table~\ref{tab:v}).

We have derived the applicability criteria for two different formulations of spin hydrodynamics. The first is based on the classical treatment of spin~\CITn{Florkowski:2018fap}, while the second uses the equilibrium Wigner function (the spin is described quantum mechanically with the help of the spin density matrix~\CITn{Bhadury:2025boe}). Interestingly, the classical and Wigner cases have the same functional form, differing only by a constant factor. The quantum approach is obtained by replacing the classical spin normalization $\spin = \sqrt{3/4}$ by~1/2. We thus see how naturally the spin value appears in the quantum framework. 

We stress that our conditions include only hydrodynamic variables and differ from those used in the past, that involved both hydrodynamic and particle properties. One of the previously used conditions had the form $\bv_*(x,p) \ll 1$ and could always be violated by particles with very large energies. 

The similarity of the results obtained for the classical and quantum cases follows from the fact that in the two calculations one has to check the large momentum behavior of the integrands that have a very similar asymptotics in this limit (both the hyperbolic sine appearing in the classical description and the hyperbolic cosine appearing in the quantum case approach the exponential function for large arguments). Because of this behavior, we expect our results to hold for Fermi-Dirac statistics as well, although this still needs to be worked out in the quantum formulation (the Fermi-Dirac case with classical spin treatment is discussed in~\CITn{Drogosz:2024gzv,Drogosz:2025ose}).

\bigskip
{\it Acknowledgments ---} We thank Rajeev Singh for bringing our attention to a poster presented at the Quark Matter 2025 conference. This work was supported in part by the Polish National Science Centre (NCN) Grant No.~2022/47/B/ST2/01372.


\appendix

\section{Proof of the upper bound on $\mathcal{Q}$}\label{sec:proofQ}

Here we prove that if all the components ($i=1,2,3$) of vectors $\bv$ and $\ev$ are constrained to lie in intervals
$-\omega_{\rm max} \leq b^i \leq \omega_{\rm max}$
and $-\omega_{\rm max} \leq e^i \leq \omega_{\rm max}$ for a certain positive $\omega_{\rm max}$, then the following inequality holds
\begin{equation}
\mathcal{Q} = \sqrt{ {\bv}^2 + {\ev}^2 + 2 |\ev \times \bv|} \leq (2 +\sqrt{2})\omega_{\rm max}.
\end{equation}
We square both sides, 
\begin{equation}
\mathcal{Q}^2 = {\bv}^2 + {\ev}^2 + 2 |\ev \times \bv| \leq (6 + 4\sqrt{2})\omega_{\rm max}^2,
\end{equation}
and calculate the second partial derivative with respect to $b^i$,
\begin{equation}\label{eq:der2b}
\f{\partial^2}{\partial (b^i)^2}\mathcal{Q}^2 = 2 + \f{2 \ev^2 \left((\ev \times \bv)^i \right)^2}{|\ev \times \bv|^3} \geq2,
\end{equation}
where $(\ev \times \bv)^i$ is the $i$-th component of the vector product. Analogously, the second partial derivative with respect to $e^i$ is
\begin{equation}\label{eq:der2e}
\f{\partial^2}{\partial (e^i)^2}\mathcal{Q}^2 = 2 + \f{2 \bv^2 \left((\ev \times \bv)^i \right)^2}{|\ev \times \bv|^3}\geq2.
\end{equation}
Therefore, if we keep five of the components constant and consider $\mathcal{Q}^2$ a one-variable function of the sixth one, this function is convex, and we can use Bauer's maximum principle to state that it attains its extremum at an end of the interval.

Repeating this argument for each of the components, we arrive at the conclusion that the global maximum must be reached at a corner of the six-dimensional cube~$[-\omega_{\rm max}, \omega_{\rm max}]^6$. At a corner, \mbox{$\bv = \omega_{\rm max} (\pm 1, \pm 1, \pm 1)$} and $\ev =\omega_{\rm max}  (\pm 1, \pm 1, \pm 1)$, from which follows \mbox{$\bv^2 = \ev^2 = 3 \, \omega_{\rm max}$}, and there are only two possible values of $|\ev \times \bv|$, namely $0$ and $2 \sqrt 2 \, \omega_{\rm max}$. Since~\mbox{$0 < 2 \sqrt 2 \, \omega_{\rm max}$}, it is the latter choice that corresponds to the global maximum of \mbox{$\mathcal{Q} = \sqrt{3+3+4\sqrt{2}}\,\omega_{\rm max} = (2 +\sqrt{2})\omega_{\rm max}$}.

Note that this simple argument, which did not require testing for joint convexity of $\mathcal{Q}^2$ (e.g., by testing for the positive semidefiniteness of the Hessian matrix) but only for convexity in each variable separately, could be applied because the domain of $\mathcal{Q}^2$ was a Cartesian product of intervals; it would have been insufficient if the domain had a more complex shape.


\begin{thebibliography}{61}%
\makeatletter
\providecommand \@ifxundefined [1]{%
 \@ifx{#1\undefined}
}%
\providecommand \@ifnum [1]{%
 \ifnum #1\expandafter \@firstoftwo
 \else \expandafter \@secondoftwo
 \fi
}%
\providecommand \@ifx [1]{%
 \ifx #1\expandafter \@firstoftwo
 \else \expandafter \@secondoftwo
 \fi
}%
\providecommand \natexlab [1]{#1}%
\providecommand \enquote  [1]{``#1''}%
\providecommand \bibnamefont  [1]{#1}%
\providecommand \bibfnamefont [1]{#1}%
\providecommand \citenamefont [1]{#1}%
\providecommand \href@noop [0]{\@secondoftwo}%
\providecommand \href [0]{\begingroup \@sanitize@url \@href}%
\providecommand \@href[1]{\@@startlink{#1}\@@href}%
\providecommand \@@href[1]{\endgroup#1\@@endlink}%
\providecommand \@sanitize@url [0]{\catcode `\\12\catcode `\$12\catcode `\&12\catcode `\#12\catcode `\^12\catcode `\_12\catcode `\%12\relax}%
\providecommand \@@startlink[1]{}%
\providecommand \@@endlink[0]{}%
\providecommand \url  [0]{\begingroup\@sanitize@url \@url }%
\providecommand \@url [1]{\endgroup\@href {#1}{\urlprefix }}%
\providecommand \urlprefix  [0]{URL }%
\providecommand \Eprint [0]{\href }%
\providecommand \doibase [0]{http://dx.doi.org/}%
\providecommand \selectlanguage [0]{\@gobble}%
\providecommand \bibinfo  [0]{\@secondoftwo}%
\providecommand \bibfield  [0]{\@secondoftwo}%
\providecommand \translation [1]{[#1]}%
\providecommand \BibitemOpen [0]{}%
\providecommand \bibitemStop [0]{}%
\providecommand \bibitemNoStop [0]{.\EOS\space}%
\providecommand \EOS [0]{\spacefactor3000\relax}%
\providecommand \BibitemShut  [1]{\csname bibitem#1\endcsname}%
\let\auto@bib@innerbib\@empty
\bibitem [{\citenamefont {Liang}\ and\ \citenamefont {Wang}(2005{\natexlab{a}})}]{Liang:2004ph}%
  \BibitemOpen
  \bibfield  {author} {\bibinfo {author} {\bibfnamefont {Z.-T.}\ \bibnamefont {Liang}}\ and\ \bibinfo {author} {\bibfnamefont {X.-N.}\ \bibnamefont {Wang}},\ }\href {\doibase 10.1103/PhysRevLett.94.102301} {\bibfield  {journal} {\bibinfo  {journal} {Phys. Rev. Lett.}\ }\textbf {\bibinfo {volume} {94}},\ \bibinfo {pages} {102301} (\bibinfo {year} {2005}{\natexlab{a}})},\ \bibinfo {note} {[Erratum: Phys.Rev.Lett. 96, 039901 (2006)]},\ \Eprint {http://arxiv.org/abs/nucl-th/0410079} {arXiv:nucl-th/0410079} \BibitemShut {NoStop}%
\bibitem [{\citenamefont {Liang}\ and\ \citenamefont {Wang}(2005{\natexlab{b}})}]{Liang:2004xn}%
  \BibitemOpen
  \bibfield  {author} {\bibinfo {author} {\bibfnamefont {Z.-T.}\ \bibnamefont {Liang}}\ and\ \bibinfo {author} {\bibfnamefont {X.-N.}\ \bibnamefont {Wang}},\ }\href {\doibase 10.1016/j.physletb.2005.09.060} {\bibfield  {journal} {\bibinfo  {journal} {Phys. Lett. B}\ }\textbf {\bibinfo {volume} {629}},\ \bibinfo {pages} {20} (\bibinfo {year} {2005}{\natexlab{b}})},\ \Eprint {http://arxiv.org/abs/nucl-th/0411101} {arXiv:nucl-th/0411101} \BibitemShut {NoStop}%
\bibitem [{\citenamefont {Adamczyk}\ \emph {et~al.}(2017)\citenamefont {Adamczyk} \emph {et~al.}}]{STAR:2017ckg}%
  \BibitemOpen
  \bibfield  {author} {\bibinfo {author} {\bibfnamefont {L.}~\bibnamefont {Adamczyk}} \emph {et~al.} (\bibinfo {collaboration} {STAR}),\ }\href {\doibase 10.1038/nature23004} {\bibfield  {journal} {\bibinfo  {journal} {Nature}\ }\textbf {\bibinfo {volume} {548}},\ \bibinfo {pages} {62} (\bibinfo {year} {2017})},\ \Eprint {http://arxiv.org/abs/1701.06657} {arXiv:1701.06657 [nucl-ex]} \BibitemShut {NoStop}%
\bibitem [{\citenamefont {Adam}\ \emph {et~al.}(2018)\citenamefont {Adam} \emph {et~al.}}]{STAR:2018gyt}%
  \BibitemOpen
  \bibfield  {author} {\bibinfo {author} {\bibfnamefont {J.}~\bibnamefont {Adam}} \emph {et~al.} (\bibinfo {collaboration} {STAR}),\ }\href {\doibase 10.1103/PhysRevC.98.014910} {\bibfield  {journal} {\bibinfo  {journal} {Phys. Rev. C}\ }\textbf {\bibinfo {volume} {98}},\ \bibinfo {pages} {014910} (\bibinfo {year} {2018})},\ \Eprint {http://arxiv.org/abs/1805.04400} {arXiv:1805.04400 [nucl-ex]} \BibitemShut {NoStop}%
\bibitem [{\citenamefont {Adam}\ \emph {et~al.}(2019)\citenamefont {Adam} \emph {et~al.}}]{STAR:2019erd}%
  \BibitemOpen
  \bibfield  {author} {\bibinfo {author} {\bibfnamefont {J.}~\bibnamefont {Adam}} \emph {et~al.} (\bibinfo {collaboration} {STAR}),\ }\href {\doibase 10.1103/PhysRevLett.123.132301} {\bibfield  {journal} {\bibinfo  {journal} {Phys. Rev. Lett.}\ }\textbf {\bibinfo {volume} {123}},\ \bibinfo {pages} {132301} (\bibinfo {year} {2019})},\ \Eprint {http://arxiv.org/abs/1905.11917} {arXiv:1905.11917 [nucl-ex]} \BibitemShut {NoStop}%
\bibitem [{\citenamefont {Acharya}\ \emph {et~al.}(2020)\citenamefont {Acharya} \emph {et~al.}}]{ALICE:2019aid}%
  \BibitemOpen
  \bibfield  {author} {\bibinfo {author} {\bibfnamefont {S.}~\bibnamefont {Acharya}} \emph {et~al.} (\bibinfo {collaboration} {ALICE}),\ }\href {\doibase 10.1103/PhysRevLett.125.012301} {\bibfield  {journal} {\bibinfo  {journal} {Phys. Rev. Lett.}\ }\textbf {\bibinfo {volume} {125}},\ \bibinfo {pages} {012301} (\bibinfo {year} {2020})},\ \Eprint {http://arxiv.org/abs/1910.14408} {arXiv:1910.14408 [nucl-ex]} \BibitemShut {NoStop}%
\bibitem [{\citenamefont {Niida}\ and\ \citenamefont {Voloshin}(2024)}]{Niida:2024ntm}%
  \BibitemOpen
  \bibfield  {author} {\bibinfo {author} {\bibfnamefont {T.}~\bibnamefont {Niida}}\ and\ \bibinfo {author} {\bibfnamefont {S.~A.}\ \bibnamefont {Voloshin}},\ }\href {\doibase 10.1142/S0218301324300108} {\bibfield  {journal} {\bibinfo  {journal} {Int. J. Mod. Phys. E}\ }\textbf {\bibinfo {volume} {33}},\ \bibinfo {pages} {2430010} (\bibinfo {year} {2024})},\ \Eprint {http://arxiv.org/abs/2404.11042} {arXiv:2404.11042 [nucl-ex]} \BibitemShut {NoStop}%
\bibitem [{\citenamefont {Becattini}\ and\ \citenamefont {Tinti}(2010)}]{Becattini:2009wh}%
  \BibitemOpen
  \bibfield  {author} {\bibinfo {author} {\bibfnamefont {F.}~\bibnamefont {Becattini}}\ and\ \bibinfo {author} {\bibfnamefont {L.}~\bibnamefont {Tinti}},\ }\href {\doibase 10.1016/j.aop.2010.03.007} {\bibfield  {journal} {\bibinfo  {journal} {Annals Phys.}\ }\textbf {\bibinfo {volume} {325}},\ \bibinfo {pages} {1566} (\bibinfo {year} {2010})},\ \Eprint {http://arxiv.org/abs/0911.0864} {arXiv:0911.0864 [gr-qc]} \BibitemShut {NoStop}%
\bibitem [{\citenamefont {Becattini}(2011)}]{Becattini:2011zz}%
  \BibitemOpen
  \bibfield  {author} {\bibinfo {author} {\bibfnamefont {F.}~\bibnamefont {Becattini}},\ }\href {\doibase 10.1134/S1547477111080036} {\bibfield  {journal} {\bibinfo  {journal} {Phys. Part. Nucl. Lett.}\ }\textbf {\bibinfo {volume} {8}},\ \bibinfo {pages} {801} (\bibinfo {year} {2011})}\BibitemShut {NoStop}%
\bibitem [{\citenamefont {Becattini}\ \emph {et~al.}(2013)\citenamefont {Becattini}, \citenamefont {Chandra}, \citenamefont {Del~Zanna},\ and\ \citenamefont {Grossi}}]{Becattini:2013fla}%
  \BibitemOpen
  \bibfield  {author} {\bibinfo {author} {\bibfnamefont {F.}~\bibnamefont {Becattini}}, \bibinfo {author} {\bibfnamefont {V.}~\bibnamefont {Chandra}}, \bibinfo {author} {\bibfnamefont {L.}~\bibnamefont {Del~Zanna}}, \ and\ \bibinfo {author} {\bibfnamefont {E.}~\bibnamefont {Grossi}},\ }\href {\doibase 10.1016/j.aop.2013.07.004} {\bibfield  {journal} {\bibinfo  {journal} {Annals Phys.}\ }\textbf {\bibinfo {volume} {338}},\ \bibinfo {pages} {32} (\bibinfo {year} {2013})},\ \Eprint {http://arxiv.org/abs/1303.3431} {arXiv:1303.3431 [nucl-th]} \BibitemShut {NoStop}%
\bibitem [{\citenamefont {Montenegro}\ \emph {et~al.}(2017)\citenamefont {Montenegro}, \citenamefont {Tinti},\ and\ \citenamefont {Torrieri}}]{Montenegro:2017rbu}%
  \BibitemOpen
  \bibfield  {author} {\bibinfo {author} {\bibfnamefont {D.}~\bibnamefont {Montenegro}}, \bibinfo {author} {\bibfnamefont {L.}~\bibnamefont {Tinti}}, \ and\ \bibinfo {author} {\bibfnamefont {G.}~\bibnamefont {Torrieri}},\ }\href {\doibase 10.1103/PhysRevD.96.056012} {\bibfield  {journal} {\bibinfo  {journal} {Phys. Rev. D}\ }\textbf {\bibinfo {volume} {96}},\ \bibinfo {pages} {056012} (\bibinfo {year} {2017})},\ \bibinfo {note} {[Addendum: Phys.Rev.D 96, 079901 (2017)]},\ \Eprint {http://arxiv.org/abs/1701.08263} {arXiv:1701.08263 [hep-th]} \BibitemShut {NoStop}%
\bibitem [{\citenamefont {Florkowski}\ \emph {et~al.}(2018{\natexlab{a}})\citenamefont {Florkowski}, \citenamefont {Friman}, \citenamefont {Jaiswal},\ and\ \citenamefont {Speranza}}]{Florkowski:2017ruc}%
  \BibitemOpen
  \bibfield  {author} {\bibinfo {author} {\bibfnamefont {W.}~\bibnamefont {Florkowski}}, \bibinfo {author} {\bibfnamefont {B.}~\bibnamefont {Friman}}, \bibinfo {author} {\bibfnamefont {A.}~\bibnamefont {Jaiswal}}, \ and\ \bibinfo {author} {\bibfnamefont {E.}~\bibnamefont {Speranza}},\ }\href {\doibase 10.1103/PhysRevC.97.041901} {\bibfield  {journal} {\bibinfo  {journal} {Phys. Rev.}\ }\textbf {\bibinfo {volume} {C97}},\ \bibinfo {pages} {041901} (\bibinfo {year} {2018}{\natexlab{a}})},\ \Eprint {http://arxiv.org/abs/1705.00587} {arXiv:1705.00587 [nucl-th]} \BibitemShut {NoStop}%
\bibitem [{\citenamefont {Florkowski}\ \emph {et~al.}(2018{\natexlab{b}})\citenamefont {Florkowski}, \citenamefont {Friman}, \citenamefont {Jaiswal}, \citenamefont {Ryblewski},\ and\ \citenamefont {Speranza}}]{Florkowski:2017dyn}%
  \BibitemOpen
  \bibfield  {author} {\bibinfo {author} {\bibfnamefont {W.}~\bibnamefont {Florkowski}}, \bibinfo {author} {\bibfnamefont {B.}~\bibnamefont {Friman}}, \bibinfo {author} {\bibfnamefont {A.}~\bibnamefont {Jaiswal}}, \bibinfo {author} {\bibfnamefont {R.}~\bibnamefont {Ryblewski}}, \ and\ \bibinfo {author} {\bibfnamefont {E.}~\bibnamefont {Speranza}},\ }\href {\doibase 10.1103/PhysRevD.97.116017} {\bibfield  {journal} {\bibinfo  {journal} {Phys. Rev. D}\ }\textbf {\bibinfo {volume} {97}},\ \bibinfo {pages} {116017} (\bibinfo {year} {2018}{\natexlab{b}})},\ \Eprint {http://arxiv.org/abs/1712.07676} {arXiv:1712.07676 [nucl-th]} \BibitemShut {NoStop}%
\bibitem [{\citenamefont {Florkowski}\ \emph {et~al.}(2018{\natexlab{c}})\citenamefont {Florkowski}, \citenamefont {Kumar},\ and\ \citenamefont {Ryblewski}}]{Florkowski:2018ahw}%
  \BibitemOpen
  \bibfield  {author} {\bibinfo {author} {\bibfnamefont {W.}~\bibnamefont {Florkowski}}, \bibinfo {author} {\bibfnamefont {A.}~\bibnamefont {Kumar}}, \ and\ \bibinfo {author} {\bibfnamefont {R.}~\bibnamefont {Ryblewski}},\ }\href {\doibase 10.1103/PhysRevC.98.044906} {\bibfield  {journal} {\bibinfo  {journal} {Phys. Rev. C}\ }\textbf {\bibinfo {volume} {98}},\ \bibinfo {pages} {044906} (\bibinfo {year} {2018}{\natexlab{c}})},\ \Eprint {http://arxiv.org/abs/1806.02616} {arXiv:1806.02616 [hep-ph]} \BibitemShut {NoStop}%
\bibitem [{\citenamefont {Hattori}\ \emph {et~al.}(2019)\citenamefont {Hattori}, \citenamefont {Hongo}, \citenamefont {Huang}, \citenamefont {Matsuo},\ and\ \citenamefont {Taya}}]{Hattori:2019lfp}%
  \BibitemOpen
  \bibfield  {author} {\bibinfo {author} {\bibfnamefont {K.}~\bibnamefont {Hattori}}, \bibinfo {author} {\bibfnamefont {M.}~\bibnamefont {Hongo}}, \bibinfo {author} {\bibfnamefont {X.-G.}\ \bibnamefont {Huang}}, \bibinfo {author} {\bibfnamefont {M.}~\bibnamefont {Matsuo}}, \ and\ \bibinfo {author} {\bibfnamefont {H.}~\bibnamefont {Taya}},\ }\href {\doibase 10.1016/j.physletb.2019.05.040} {\bibfield  {journal} {\bibinfo  {journal} {Phys. Lett. B}\ }\textbf {\bibinfo {volume} {795}},\ \bibinfo {pages} {100} (\bibinfo {year} {2019})},\ \Eprint {http://arxiv.org/abs/1901.06615} {arXiv:1901.06615 [hep-th]} \BibitemShut {NoStop}%
\bibitem [{\citenamefont {Weickgenannt}\ \emph {et~al.}(2019)\citenamefont {Weickgenannt}, \citenamefont {Sheng}, \citenamefont {Speranza}, \citenamefont {Wang},\ and\ \citenamefont {Rischke}}]{Weickgenannt:2019dks}%
  \BibitemOpen
  \bibfield  {author} {\bibinfo {author} {\bibfnamefont {N.}~\bibnamefont {Weickgenannt}}, \bibinfo {author} {\bibfnamefont {X.-L.}\ \bibnamefont {Sheng}}, \bibinfo {author} {\bibfnamefont {E.}~\bibnamefont {Speranza}}, \bibinfo {author} {\bibfnamefont {Q.}~\bibnamefont {Wang}}, \ and\ \bibinfo {author} {\bibfnamefont {D.~H.}\ \bibnamefont {Rischke}},\ }\href {\doibase 10.1103/PhysRevD.100.056018} {\bibfield  {journal} {\bibinfo  {journal} {Phys. Rev. D}\ }\textbf {\bibinfo {volume} {100}},\ \bibinfo {pages} {056018} (\bibinfo {year} {2019})},\ \Eprint {http://arxiv.org/abs/1902.06513} {arXiv:1902.06513 [hep-ph]} \BibitemShut {NoStop}%
\bibitem [{\citenamefont {Florkowski}\ \emph {et~al.}(2019{\natexlab{a}})\citenamefont {Florkowski}, \citenamefont {Kumar}, \citenamefont {Ryblewski},\ and\ \citenamefont {Singh}}]{Florkowski:2019qdp}%
  \BibitemOpen
  \bibfield  {author} {\bibinfo {author} {\bibfnamefont {W.}~\bibnamefont {Florkowski}}, \bibinfo {author} {\bibfnamefont {A.}~\bibnamefont {Kumar}}, \bibinfo {author} {\bibfnamefont {R.}~\bibnamefont {Ryblewski}}, \ and\ \bibinfo {author} {\bibfnamefont {R.}~\bibnamefont {Singh}},\ }\href {\doibase 10.1103/PhysRevC.99.044910} {\bibfield  {journal} {\bibinfo  {journal} {Phys. Rev. C}\ }\textbf {\bibinfo {volume} {99}},\ \bibinfo {pages} {044910} (\bibinfo {year} {2019}{\natexlab{a}})},\ \Eprint {http://arxiv.org/abs/1901.09655} {arXiv:1901.09655 [hep-ph]} \BibitemShut {NoStop}%
\bibitem [{\citenamefont {Bhadury}\ \emph {et~al.}(2021{\natexlab{a}})\citenamefont {Bhadury}, \citenamefont {Florkowski}, \citenamefont {Jaiswal}, \citenamefont {Kumar},\ and\ \citenamefont {Ryblewski}}]{Bhadury:2020puc}%
  \BibitemOpen
  \bibfield  {author} {\bibinfo {author} {\bibfnamefont {S.}~\bibnamefont {Bhadury}}, \bibinfo {author} {\bibfnamefont {W.}~\bibnamefont {Florkowski}}, \bibinfo {author} {\bibfnamefont {A.}~\bibnamefont {Jaiswal}}, \bibinfo {author} {\bibfnamefont {A.}~\bibnamefont {Kumar}}, \ and\ \bibinfo {author} {\bibfnamefont {R.}~\bibnamefont {Ryblewski}},\ }\href {\doibase 10.1016/j.physletb.2021.136096} {\bibfield  {journal} {\bibinfo  {journal} {Phys. Lett. B}\ }\textbf {\bibinfo {volume} {814}},\ \bibinfo {pages} {136096} (\bibinfo {year} {2021}{\natexlab{a}})},\ \Eprint {http://arxiv.org/abs/2002.03937} {arXiv:2002.03937 [hep-ph]} \BibitemShut {NoStop}%
\bibitem [{\citenamefont {Montenegro}\ and\ \citenamefont {Torrieri}(2020)}]{Montenegro:2020paq}%
  \BibitemOpen
  \bibfield  {author} {\bibinfo {author} {\bibfnamefont {D.}~\bibnamefont {Montenegro}}\ and\ \bibinfo {author} {\bibfnamefont {G.}~\bibnamefont {Torrieri}},\ }\href {\doibase 10.1103/PhysRevD.102.036007} {\bibfield  {journal} {\bibinfo  {journal} {Phys. Rev. D}\ }\textbf {\bibinfo {volume} {102}},\ \bibinfo {pages} {036007} (\bibinfo {year} {2020})},\ \Eprint {http://arxiv.org/abs/2004.10195} {arXiv:2004.10195 [hep-th]} \BibitemShut {NoStop}%
\bibitem [{\citenamefont {Weickgenannt}\ \emph {et~al.}(2021{\natexlab{a}})\citenamefont {Weickgenannt}, \citenamefont {Speranza}, \citenamefont {Sheng}, \citenamefont {Wang},\ and\ \citenamefont {Rischke}}]{Weickgenannt:2020aaf}%
  \BibitemOpen
  \bibfield  {author} {\bibinfo {author} {\bibfnamefont {N.}~\bibnamefont {Weickgenannt}}, \bibinfo {author} {\bibfnamefont {E.}~\bibnamefont {Speranza}}, \bibinfo {author} {\bibfnamefont {X.-l.}\ \bibnamefont {Sheng}}, \bibinfo {author} {\bibfnamefont {Q.}~\bibnamefont {Wang}}, \ and\ \bibinfo {author} {\bibfnamefont {D.~H.}\ \bibnamefont {Rischke}},\ }\href {\doibase 10.1103/PhysRevLett.127.052301} {\bibfield  {journal} {\bibinfo  {journal} {Phys. Rev. Lett.}\ }\textbf {\bibinfo {volume} {127}},\ \bibinfo {pages} {052301} (\bibinfo {year} {2021}{\natexlab{a}})},\ \Eprint {http://arxiv.org/abs/2005.01506} {arXiv:2005.01506 [hep-ph]} \BibitemShut {NoStop}%
\bibitem [{\citenamefont {Shi}\ \emph {et~al.}(2021)\citenamefont {Shi}, \citenamefont {Gale},\ and\ \citenamefont {Jeon}}]{Shi:2020htn}%
  \BibitemOpen
  \bibfield  {author} {\bibinfo {author} {\bibfnamefont {S.}~\bibnamefont {Shi}}, \bibinfo {author} {\bibfnamefont {C.}~\bibnamefont {Gale}}, \ and\ \bibinfo {author} {\bibfnamefont {S.}~\bibnamefont {Jeon}},\ }\href {\doibase 10.1103/PhysRevC.103.044906} {\bibfield  {journal} {\bibinfo  {journal} {Phys. Rev. C}\ }\textbf {\bibinfo {volume} {103}},\ \bibinfo {pages} {044906} (\bibinfo {year} {2021})},\ \Eprint {http://arxiv.org/abs/2008.08618} {arXiv:2008.08618 [nucl-th]} \BibitemShut {NoStop}%
\bibitem [{\citenamefont {Bhadury}\ \emph {et~al.}(2021{\natexlab{b}})\citenamefont {Bhadury}, \citenamefont {Florkowski}, \citenamefont {Jaiswal}, \citenamefont {Kumar},\ and\ \citenamefont {Ryblewski}}]{Bhadury:2020cop}%
  \BibitemOpen
  \bibfield  {author} {\bibinfo {author} {\bibfnamefont {S.}~\bibnamefont {Bhadury}}, \bibinfo {author} {\bibfnamefont {W.}~\bibnamefont {Florkowski}}, \bibinfo {author} {\bibfnamefont {A.}~\bibnamefont {Jaiswal}}, \bibinfo {author} {\bibfnamefont {A.}~\bibnamefont {Kumar}}, \ and\ \bibinfo {author} {\bibfnamefont {R.}~\bibnamefont {Ryblewski}},\ }\href {\doibase 10.1103/PhysRevD.103.014030} {\bibfield  {journal} {\bibinfo  {journal} {Phys. Rev. D}\ }\textbf {\bibinfo {volume} {103}},\ \bibinfo {pages} {014030} (\bibinfo {year} {2021}{\natexlab{b}})},\ \Eprint {http://arxiv.org/abs/2008.10976} {arXiv:2008.10976 [nucl-th]} \BibitemShut {NoStop}%
\bibitem [{\citenamefont {Fukushima}\ and\ \citenamefont {Pu}(2021)}]{Fukushima:2020ucl}%
  \BibitemOpen
  \bibfield  {author} {\bibinfo {author} {\bibfnamefont {K.}~\bibnamefont {Fukushima}}\ and\ \bibinfo {author} {\bibfnamefont {S.}~\bibnamefont {Pu}},\ }\href {\doibase 10.1016/j.physletb.2021.136346} {\bibfield  {journal} {\bibinfo  {journal} {Phys. Lett. B}\ }\textbf {\bibinfo {volume} {817}},\ \bibinfo {pages} {136346} (\bibinfo {year} {2021})},\ \Eprint {http://arxiv.org/abs/2010.01608} {arXiv:2010.01608 [hep-th]} \BibitemShut {NoStop}%
\bibitem [{\citenamefont {Li}\ \emph {et~al.}(2021)\citenamefont {Li}, \citenamefont {Stephanov},\ and\ \citenamefont {Yee}}]{Li:2020eon}%
  \BibitemOpen
  \bibfield  {author} {\bibinfo {author} {\bibfnamefont {S.}~\bibnamefont {Li}}, \bibinfo {author} {\bibfnamefont {M.~A.}\ \bibnamefont {Stephanov}}, \ and\ \bibinfo {author} {\bibfnamefont {H.-U.}\ \bibnamefont {Yee}},\ }\href {\doibase 10.1103/PhysRevLett.127.082302} {\bibfield  {journal} {\bibinfo  {journal} {Phys. Rev. Lett.}\ }\textbf {\bibinfo {volume} {127}},\ \bibinfo {pages} {082302} (\bibinfo {year} {2021})},\ \Eprint {http://arxiv.org/abs/2011.12318} {arXiv:2011.12318 [hep-th]} \BibitemShut {NoStop}%
\bibitem [{\citenamefont {Singh}\ \emph {et~al.}(2021)\citenamefont {Singh}, \citenamefont {Sophys},\ and\ \citenamefont {Ryblewski}}]{Singh:2020rht}%
  \BibitemOpen
  \bibfield  {author} {\bibinfo {author} {\bibfnamefont {R.}~\bibnamefont {Singh}}, \bibinfo {author} {\bibfnamefont {G.}~\bibnamefont {Sophys}}, \ and\ \bibinfo {author} {\bibfnamefont {R.}~\bibnamefont {Ryblewski}},\ }\href {\doibase 10.1103/PhysRevD.103.074024} {\bibfield  {journal} {\bibinfo  {journal} {Phys. Rev. D}\ }\textbf {\bibinfo {volume} {103}},\ \bibinfo {pages} {074024} (\bibinfo {year} {2021})},\ \Eprint {http://arxiv.org/abs/2011.14907} {arXiv:2011.14907 [hep-ph]} \BibitemShut {NoStop}%
\bibitem [{\citenamefont {Gallegos}\ \emph {et~al.}(2021)\citenamefont {Gallegos}, \citenamefont {G\"ursoy},\ and\ \citenamefont {Yarom}}]{Gallegos:2021bzp}%
  \BibitemOpen
  \bibfield  {author} {\bibinfo {author} {\bibfnamefont {A.~D.}\ \bibnamefont {Gallegos}}, \bibinfo {author} {\bibfnamefont {U.}~\bibnamefont {G\"ursoy}}, \ and\ \bibinfo {author} {\bibfnamefont {A.}~\bibnamefont {Yarom}},\ }\href {\doibase 10.21468/SciPostPhys.11.2.041} {\bibfield  {journal} {\bibinfo  {journal} {SciPost Phys.}\ }\textbf {\bibinfo {volume} {11}},\ \bibinfo {pages} {041} (\bibinfo {year} {2021})},\ \Eprint {http://arxiv.org/abs/2101.04759} {arXiv:2101.04759 [hep-th]} \BibitemShut {NoStop}%
\bibitem [{\citenamefont {Weickgenannt}\ \emph {et~al.}(2021{\natexlab{b}})\citenamefont {Weickgenannt}, \citenamefont {Speranza}, \citenamefont {Sheng}, \citenamefont {Wang},\ and\ \citenamefont {Rischke}}]{Weickgenannt:2021cuo}%
  \BibitemOpen
  \bibfield  {author} {\bibinfo {author} {\bibfnamefont {N.}~\bibnamefont {Weickgenannt}}, \bibinfo {author} {\bibfnamefont {E.}~\bibnamefont {Speranza}}, \bibinfo {author} {\bibfnamefont {X.-l.}\ \bibnamefont {Sheng}}, \bibinfo {author} {\bibfnamefont {Q.}~\bibnamefont {Wang}}, \ and\ \bibinfo {author} {\bibfnamefont {D.~H.}\ \bibnamefont {Rischke}},\ }\href {\doibase 10.1103/PhysRevD.104.016022} {\bibfield  {journal} {\bibinfo  {journal} {Phys. Rev. D}\ }\textbf {\bibinfo {volume} {104}},\ \bibinfo {pages} {016022} (\bibinfo {year} {2021}{\natexlab{b}})},\ \Eprint {http://arxiv.org/abs/2103.04896} {arXiv:2103.04896 [nucl-th]} \BibitemShut {NoStop}%
\bibitem [{\citenamefont {She}\ \emph {et~al.}(2022)\citenamefont {She}, \citenamefont {Huang}, \citenamefont {Hou},\ and\ \citenamefont {Liao}}]{She:2021lhe}%
  \BibitemOpen
  \bibfield  {author} {\bibinfo {author} {\bibfnamefont {D.}~\bibnamefont {She}}, \bibinfo {author} {\bibfnamefont {A.}~\bibnamefont {Huang}}, \bibinfo {author} {\bibfnamefont {D.}~\bibnamefont {Hou}}, \ and\ \bibinfo {author} {\bibfnamefont {J.}~\bibnamefont {Liao}},\ }\href {\doibase 10.1016/j.scib.2022.10.020} {\bibfield  {journal} {\bibinfo  {journal} {Sci. Bull.}\ }\textbf {\bibinfo {volume} {67}},\ \bibinfo {pages} {2265} (\bibinfo {year} {2022})},\ \Eprint {http://arxiv.org/abs/2105.04060} {arXiv:2105.04060 [nucl-th]} \BibitemShut {NoStop}%
\bibitem [{\citenamefont {Hongo}\ \emph {et~al.}(2021)\citenamefont {Hongo}, \citenamefont {Huang}, \citenamefont {Kaminski}, \citenamefont {Stephanov},\ and\ \citenamefont {Yee}}]{Hongo:2021ona}%
  \BibitemOpen
  \bibfield  {author} {\bibinfo {author} {\bibfnamefont {M.}~\bibnamefont {Hongo}}, \bibinfo {author} {\bibfnamefont {X.-G.}\ \bibnamefont {Huang}}, \bibinfo {author} {\bibfnamefont {M.}~\bibnamefont {Kaminski}}, \bibinfo {author} {\bibfnamefont {M.}~\bibnamefont {Stephanov}}, \ and\ \bibinfo {author} {\bibfnamefont {H.-U.}\ \bibnamefont {Yee}},\ }\href {\doibase 10.1007/JHEP11(2021)150} {\bibfield  {journal} {\bibinfo  {journal} {JHEP}\ }\textbf {\bibinfo {volume} {11}},\ \bibinfo {pages} {150} (\bibinfo {year} {2021})},\ \Eprint {http://arxiv.org/abs/2107.14231} {arXiv:2107.14231 [hep-th]} \BibitemShut {NoStop}%
\bibitem [{\citenamefont {Hu}(2022)}]{Hu:2021pwh}%
  \BibitemOpen
  \bibfield  {author} {\bibinfo {author} {\bibfnamefont {J.}~\bibnamefont {Hu}},\ }\href {\doibase 10.1103/PhysRevD.105.076009} {\bibfield  {journal} {\bibinfo  {journal} {Phys. Rev. D}\ }\textbf {\bibinfo {volume} {105}},\ \bibinfo {pages} {076009} (\bibinfo {year} {2022})},\ \Eprint {http://arxiv.org/abs/2111.03571} {arXiv:2111.03571 [hep-ph]} \BibitemShut {NoStop}%
\bibitem [{\citenamefont {Singh}\ \emph {et~al.}(2023)\citenamefont {Singh}, \citenamefont {Shokri},\ and\ \citenamefont {Mehr}}]{Singh:2022ltu}%
  \BibitemOpen
  \bibfield  {author} {\bibinfo {author} {\bibfnamefont {R.}~\bibnamefont {Singh}}, \bibinfo {author} {\bibfnamefont {M.}~\bibnamefont {Shokri}}, \ and\ \bibinfo {author} {\bibfnamefont {S.~M. A.~T.}\ \bibnamefont {Mehr}},\ }\href {\doibase 10.1016/j.nuclphysa.2023.122656} {\bibfield  {journal} {\bibinfo  {journal} {Nucl. Phys. A}\ }\textbf {\bibinfo {volume} {1035}},\ \bibinfo {pages} {122656} (\bibinfo {year} {2023})},\ \Eprint {http://arxiv.org/abs/2202.11504} {arXiv:2202.11504 [hep-ph]} \BibitemShut {NoStop}%
\bibitem [{\citenamefont {Weickgenannt}\ \emph {et~al.}(2022)\citenamefont {Weickgenannt}, \citenamefont {Wagner}, \citenamefont {Speranza},\ and\ \citenamefont {Rischke}}]{Weickgenannt:2022zxs}%
  \BibitemOpen
  \bibfield  {author} {\bibinfo {author} {\bibfnamefont {N.}~\bibnamefont {Weickgenannt}}, \bibinfo {author} {\bibfnamefont {D.}~\bibnamefont {Wagner}}, \bibinfo {author} {\bibfnamefont {E.}~\bibnamefont {Speranza}}, \ and\ \bibinfo {author} {\bibfnamefont {D.~H.}\ \bibnamefont {Rischke}},\ }\href {\doibase 10.1103/PhysRevD.106.096014} {\bibfield  {journal} {\bibinfo  {journal} {Phys. Rev. D}\ }\textbf {\bibinfo {volume} {106}},\ \bibinfo {pages} {096014} (\bibinfo {year} {2022})},\ \Eprint {http://arxiv.org/abs/2203.04766} {arXiv:2203.04766 [nucl-th]} \BibitemShut {NoStop}%
\bibitem [{\citenamefont {Wagner}\ \emph {et~al.}(2022)\citenamefont {Wagner}, \citenamefont {Weickgenannt},\ and\ \citenamefont {Rischke}}]{Wagner:2022amr}%
  \BibitemOpen
  \bibfield  {author} {\bibinfo {author} {\bibfnamefont {D.}~\bibnamefont {Wagner}}, \bibinfo {author} {\bibfnamefont {N.}~\bibnamefont {Weickgenannt}}, \ and\ \bibinfo {author} {\bibfnamefont {D.~H.}\ \bibnamefont {Rischke}},\ }\href {\doibase 10.1103/PhysRevD.106.116021} {\bibfield  {journal} {\bibinfo  {journal} {Phys. Rev. D}\ }\textbf {\bibinfo {volume} {106}},\ \bibinfo {pages} {116021} (\bibinfo {year} {2022})},\ \Eprint {http://arxiv.org/abs/2210.06187} {arXiv:2210.06187 [nucl-th]} \BibitemShut {NoStop}%
\bibitem [{\citenamefont {Dey}\ \emph {et~al.}(2023)\citenamefont {Dey}, \citenamefont {Florkowski}, \citenamefont {Jaiswal},\ and\ \citenamefont {Ryblewski}}]{Dey:2023hft}%
  \BibitemOpen
  \bibfield  {author} {\bibinfo {author} {\bibfnamefont {S.}~\bibnamefont {Dey}}, \bibinfo {author} {\bibfnamefont {W.}~\bibnamefont {Florkowski}}, \bibinfo {author} {\bibfnamefont {A.}~\bibnamefont {Jaiswal}}, \ and\ \bibinfo {author} {\bibfnamefont {R.}~\bibnamefont {Ryblewski}},\ }\href {\doibase 10.1016/j.physletb.2023.137994} {\bibfield  {journal} {\bibinfo  {journal} {Phys. Lett. B}\ }\textbf {\bibinfo {volume} {843}},\ \bibinfo {pages} {137994} (\bibinfo {year} {2023})},\ \Eprint {http://arxiv.org/abs/2303.05271} {arXiv:2303.05271 [hep-th]} \BibitemShut {NoStop}%
\bibitem [{\citenamefont {Weickgenannt}\ and\ \citenamefont {Blaizot}(2024)}]{Weickgenannt:2023nge}%
  \BibitemOpen
  \bibfield  {author} {\bibinfo {author} {\bibfnamefont {N.}~\bibnamefont {Weickgenannt}}\ and\ \bibinfo {author} {\bibfnamefont {J.-P.}\ \bibnamefont {Blaizot}},\ }\href {\doibase 10.1103/PhysRevD.109.056012} {\bibfield  {journal} {\bibinfo  {journal} {Phys. Rev. D}\ }\textbf {\bibinfo {volume} {109}},\ \bibinfo {pages} {056012} (\bibinfo {year} {2024})},\ \Eprint {http://arxiv.org/abs/2311.15817} {arXiv:2311.15817 [hep-ph]} \BibitemShut {NoStop}%
\bibitem [{\citenamefont {Kumar}\ \emph {et~al.}(2024)\citenamefont {Kumar}, \citenamefont {Yang},\ and\ \citenamefont {Gubler}}]{Kumar:2023ojl}%
  \BibitemOpen
  \bibfield  {author} {\bibinfo {author} {\bibfnamefont {A.}~\bibnamefont {Kumar}}, \bibinfo {author} {\bibfnamefont {D.-L.}\ \bibnamefont {Yang}}, \ and\ \bibinfo {author} {\bibfnamefont {P.}~\bibnamefont {Gubler}},\ }\href {\doibase 10.1103/PhysRevD.109.054038} {\bibfield  {journal} {\bibinfo  {journal} {Phys. Rev. D}\ }\textbf {\bibinfo {volume} {109}},\ \bibinfo {pages} {054038} (\bibinfo {year} {2024})},\ \Eprint {http://arxiv.org/abs/2312.16900} {arXiv:2312.16900 [nucl-th]} \BibitemShut {NoStop}%
\bibitem [{\citenamefont {Wagner}\ \emph {et~al.}(2024)\citenamefont {Wagner}, \citenamefont {Shokri},\ and\ \citenamefont {Rischke}}]{Wagner:2024fhf}%
  \BibitemOpen
  \bibfield  {author} {\bibinfo {author} {\bibfnamefont {D.}~\bibnamefont {Wagner}}, \bibinfo {author} {\bibfnamefont {M.}~\bibnamefont {Shokri}}, \ and\ \bibinfo {author} {\bibfnamefont {D.~H.}\ \bibnamefont {Rischke}},\ }\href {\doibase 10.1103/PhysRevResearch.6.043103} {\bibfield  {journal} {\bibinfo  {journal} {Phys. Rev. Res.}\ }\textbf {\bibinfo {volume} {6}},\ \bibinfo {pages} {043103} (\bibinfo {year} {2024})},\ \Eprint {http://arxiv.org/abs/2405.00533} {arXiv:2405.00533 [nucl-th]} \BibitemShut {NoStop}%
\bibitem [{\citenamefont {Wagner}(2025)}]{Wagner:2024fry}%
  \BibitemOpen
  \bibfield  {author} {\bibinfo {author} {\bibfnamefont {D.}~\bibnamefont {Wagner}},\ }\href {\doibase 10.1103/PhysRevD.111.016008} {\bibfield  {journal} {\bibinfo  {journal} {Phys. Rev. D}\ }\textbf {\bibinfo {volume} {111}},\ \bibinfo {pages} {016008} (\bibinfo {year} {2025})},\ \Eprint {http://arxiv.org/abs/2409.07143} {arXiv:2409.07143 [nucl-th]} \BibitemShut {NoStop}%
\bibitem [{\citenamefont {Dey}\ and\ \citenamefont {Das}(2025)}]{Dey:2024cwo}%
  \BibitemOpen
  \bibfield  {author} {\bibinfo {author} {\bibfnamefont {S.}~\bibnamefont {Dey}}\ and\ \bibinfo {author} {\bibfnamefont {A.}~\bibnamefont {Das}},\ }\href {\doibase 10.1103/PhysRevD.111.074037} {\bibfield  {journal} {\bibinfo  {journal} {Phys. Rev. D}\ }\textbf {\bibinfo {volume} {111}},\ \bibinfo {pages} {074037} (\bibinfo {year} {2025})},\ \Eprint {http://arxiv.org/abs/2410.04141} {arXiv:2410.04141 [nucl-th]} \BibitemShut {NoStop}%
\bibitem [{\citenamefont {She}\ \emph {et~al.}(2025)\citenamefont {She}, \citenamefont {Qiu},\ and\ \citenamefont {Hou}}]{She:2024rnx}%
  \BibitemOpen
  \bibfield  {author} {\bibinfo {author} {\bibfnamefont {D.}~\bibnamefont {She}}, \bibinfo {author} {\bibfnamefont {Y.-W.}\ \bibnamefont {Qiu}}, \ and\ \bibinfo {author} {\bibfnamefont {D.}~\bibnamefont {Hou}},\ }\href {\doibase 10.1103/PhysRevD.111.036027} {\bibfield  {journal} {\bibinfo  {journal} {Phys. Rev. D}\ }\textbf {\bibinfo {volume} {111}},\ \bibinfo {pages} {036027} (\bibinfo {year} {2025})},\ \Eprint {http://arxiv.org/abs/2410.15142} {arXiv:2410.15142 [nucl-th]} \BibitemShut {NoStop}%
\bibitem [{\citenamefont {Huang}(2024)}]{Huang:2024ffg}%
  \BibitemOpen
  \bibfield  {author} {\bibinfo {author} {\bibfnamefont {X.-G.}\ \bibnamefont {Huang}},\ }\href@noop {} {\  (\bibinfo {year} {2024})},\ \Eprint {http://arxiv.org/abs/2411.11753} {arXiv:2411.11753 [nucl-th]} \BibitemShut {NoStop}%
\bibitem [{\citenamefont {Bhadury}(2025)}]{Bhadury:2025fil}%
  \BibitemOpen
  \bibfield  {author} {\bibinfo {author} {\bibfnamefont {S.}~\bibnamefont {Bhadury}},\ }\href@noop {} {\  (\bibinfo {year} {2025})},\ \Eprint {http://arxiv.org/abs/2503.08428} {arXiv:2503.08428 [hep-ph]} \BibitemShut {NoStop}%
\bibitem [{\citenamefont {Dey}(2025)}]{Dey:2025wqw}%
  \BibitemOpen
  \bibfield  {author} {\bibinfo {author} {\bibfnamefont {S.}~\bibnamefont {Dey}},\ }\href@noop {} {\  (\bibinfo {year} {2025})},\ \Eprint {http://arxiv.org/abs/2504.18388} {arXiv:2504.18388 [hep-th]} \BibitemShut {NoStop}%
\bibitem [{\citenamefont {Weickgenannt}\ and\ \citenamefont {Blaizot}(2025)}]{Weickgenannt:2024ibf}%
  \BibitemOpen
  \bibfield  {author} {\bibinfo {author} {\bibfnamefont {N.}~\bibnamefont {Weickgenannt}}\ and\ \bibinfo {author} {\bibfnamefont {J.-P.}\ \bibnamefont {Blaizot}},\ }\href {\doibase 10.1103/PhysRevD.111.056006} {\bibfield  {journal} {\bibinfo  {journal} {Phys. Rev. D}\ }\textbf {\bibinfo {volume} {111}},\ \bibinfo {pages} {056006} (\bibinfo {year} {2025})},\ \Eprint {http://arxiv.org/abs/2409.11045} {arXiv:2409.11045 [hep-ph]} \BibitemShut {NoStop}%
\bibitem [{\citenamefont {Ollitrault}(2008)}]{Ollitrault:2007du}%
  \BibitemOpen
  \bibfield  {author} {\bibinfo {author} {\bibfnamefont {J.-Y.}\ \bibnamefont {Ollitrault}},\ }\href {\doibase 10.1088/0143-0807/29/2/010} {\bibfield  {journal} {\bibinfo  {journal} {Eur. J. Phys.}\ }\textbf {\bibinfo {volume} {29}},\ \bibinfo {pages} {275} (\bibinfo {year} {2008})},\ \Eprint {http://arxiv.org/abs/0708.2433} {arXiv:0708.2433 [nucl-th]} \BibitemShut {NoStop}%
\bibitem [{\citenamefont {Romatschke}(2010)}]{Romatschke:2009im}%
  \BibitemOpen
  \bibfield  {author} {\bibinfo {author} {\bibfnamefont {P.}~\bibnamefont {Romatschke}},\ }\href {\doibase 10.1142/S0218301310014613} {\bibfield  {journal} {\bibinfo  {journal} {Int. J. Mod. Phys. E}\ }\textbf {\bibinfo {volume} {19}},\ \bibinfo {pages} {1} (\bibinfo {year} {2010})},\ \Eprint {http://arxiv.org/abs/0902.3663} {arXiv:0902.3663 [hep-ph]} \BibitemShut {NoStop}%
\bibitem [{\citenamefont {Florkowski}(2010)}]{Florkowski:2010zz}%
  \BibitemOpen
  \bibfield  {author} {\bibinfo {author} {\bibfnamefont {W.}~\bibnamefont {Florkowski}},\ }\href@noop {} {\emph {\bibinfo {title} {{Phenomenology of Ultra-Relativistic Heavy-Ion Collisions}}}}\ (\bibinfo  {publisher} {Singapore: World Scientific},\ \bibinfo {year} {2010})\BibitemShut {NoStop}%
\bibitem [{\citenamefont {Gale}\ \emph {et~al.}(2013)\citenamefont {Gale}, \citenamefont {Jeon},\ and\ \citenamefont {Schenke}}]{Gale:2013da}%
  \BibitemOpen
  \bibfield  {author} {\bibinfo {author} {\bibfnamefont {C.}~\bibnamefont {Gale}}, \bibinfo {author} {\bibfnamefont {S.}~\bibnamefont {Jeon}}, \ and\ \bibinfo {author} {\bibfnamefont {B.}~\bibnamefont {Schenke}},\ }\href {\doibase 10.1142/S0217751X13400113} {\bibfield  {journal} {\bibinfo  {journal} {Int. J. Mod. Phys. A}\ }\textbf {\bibinfo {volume} {28}},\ \bibinfo {pages} {1340011} (\bibinfo {year} {2013})},\ \Eprint {http://arxiv.org/abs/1301.5893} {arXiv:1301.5893 [nucl-th]} \BibitemShut {NoStop}%
\bibitem [{\citenamefont {Jaiswal}\ and\ \citenamefont {Roy}(2016)}]{Jaiswal:2016hex}%
  \BibitemOpen
  \bibfield  {author} {\bibinfo {author} {\bibfnamefont {A.}~\bibnamefont {Jaiswal}}\ and\ \bibinfo {author} {\bibfnamefont {V.}~\bibnamefont {Roy}},\ }\href {\doibase 10.1155/2016/9623034} {\bibfield  {journal} {\bibinfo  {journal} {Adv. High Energy Phys.}\ }\textbf {\bibinfo {volume} {2016}},\ \bibinfo {pages} {9623034} (\bibinfo {year} {2016})},\ \Eprint {http://arxiv.org/abs/1605.08694} {arXiv:1605.08694 [nucl-th]} \BibitemShut {NoStop}%
\bibitem [{\citenamefont {Singh}\ \emph {et~al.}(2025)\citenamefont {Singh}, \citenamefont {Ryblewski},\ and\ \citenamefont {Florkowski}}]{Singh:2024cub}%
  \BibitemOpen
  \bibfield  {author} {\bibinfo {author} {\bibfnamefont {S.~K.}\ \bibnamefont {Singh}}, \bibinfo {author} {\bibfnamefont {R.}~\bibnamefont {Ryblewski}}, \ and\ \bibinfo {author} {\bibfnamefont {W.}~\bibnamefont {Florkowski}},\ }\href {\doibase 10.1103/PhysRevC.111.024907} {\bibfield  {journal} {\bibinfo  {journal} {Phys. Rev. C}\ }\textbf {\bibinfo {volume} {111}},\ \bibinfo {pages} {024907} (\bibinfo {year} {2025})},\ \Eprint {http://arxiv.org/abs/2411.08223} {arXiv:2411.08223 [hep-ph]} \BibitemShut {NoStop}%
\bibitem [{\citenamefont {Sapna}\ \emph {et~al.}(2025)\citenamefont {Sapna}, \citenamefont {Singh},\ and\ \citenamefont {Wagner}}]{Sapna:2025yss}%
  \BibitemOpen
  \bibfield  {author} {\bibinfo {author} {\bibnamefont {Sapna}}, \bibinfo {author} {\bibfnamefont {S.~K.}\ \bibnamefont {Singh}}, \ and\ \bibinfo {author} {\bibfnamefont {D.}~\bibnamefont {Wagner}},\ }\href@noop {} {\  (\bibinfo {year} {2025})},\ \Eprint {http://arxiv.org/abs/2503.22552} {arXiv:2503.22552 [hep-ph]} \BibitemShut {NoStop}%
\bibitem [{\citenamefont {Florkowski}\ \emph {et~al.}(2019{\natexlab{b}})\citenamefont {Florkowski}, \citenamefont {Kumar},\ and\ \citenamefont {Ryblewski}}]{Florkowski:2018fap}%
  \BibitemOpen
  \bibfield  {author} {\bibinfo {author} {\bibfnamefont {W.}~\bibnamefont {Florkowski}}, \bibinfo {author} {\bibfnamefont {A.}~\bibnamefont {Kumar}}, \ and\ \bibinfo {author} {\bibfnamefont {R.}~\bibnamefont {Ryblewski}},\ }\href {\doibase 10.1016/j.ppnp.2019.07.001} {\bibfield  {journal} {\bibinfo  {journal} {Prog. Part. Nucl. Phys.}\ }\textbf {\bibinfo {volume} {108}},\ \bibinfo {pages} {103709} (\bibinfo {year} {2019}{\natexlab{b}})},\ \Eprint {http://arxiv.org/abs/1811.04409} {arXiv:1811.04409 [nucl-th]} \BibitemShut {NoStop}%
\bibitem [{\citenamefont {Florkowski}\ and\ \citenamefont {Hontarenko}(2025)}]{Florkowski:2024bfw}%
  \BibitemOpen
  \bibfield  {author} {\bibinfo {author} {\bibfnamefont {W.}~\bibnamefont {Florkowski}}\ and\ \bibinfo {author} {\bibfnamefont {M.}~\bibnamefont {Hontarenko}},\ }\href {\doibase 10.1103/PhysRevLett.134.082302} {\bibfield  {journal} {\bibinfo  {journal} {Phys. Rev. Lett.}\ }\textbf {\bibinfo {volume} {134}},\ \bibinfo {pages} {082302} (\bibinfo {year} {2025})},\ \Eprint {http://arxiv.org/abs/2405.03263} {arXiv:2405.03263 [hep-ph]} \BibitemShut {NoStop}%
\bibitem [{\citenamefont {Drogosz}\ \emph {et~al.}(2024)\citenamefont {Drogosz}, \citenamefont {Florkowski},\ and\ \citenamefont {Hontarenko}}]{Drogosz:2024gzv}%
  \BibitemOpen
  \bibfield  {author} {\bibinfo {author} {\bibfnamefont {Z.}~\bibnamefont {Drogosz}}, \bibinfo {author} {\bibfnamefont {W.}~\bibnamefont {Florkowski}}, \ and\ \bibinfo {author} {\bibfnamefont {M.}~\bibnamefont {Hontarenko}},\ }\href {\doibase 10.1103/PhysRevD.110.096018} {\bibfield  {journal} {\bibinfo  {journal} {Phys. Rev. D}\ }\textbf {\bibinfo {volume} {110}},\ \bibinfo {pages} {096018} (\bibinfo {year} {2024})},\ \Eprint {http://arxiv.org/abs/2408.03106} {arXiv:2408.03106 [hep-ph]} \BibitemShut {NoStop}%
\bibitem [{\citenamefont {Bhadury}\ \emph {et~al.}(2025)\citenamefont {Bhadury}, \citenamefont {Drogosz}, \citenamefont {Florkowski}, \citenamefont {Kar},\ and\ \citenamefont {Mykhaylova}}]{Bhadury:2025boe}%
  \BibitemOpen
  \bibfield  {author} {\bibinfo {author} {\bibfnamefont {S.}~\bibnamefont {Bhadury}}, \bibinfo {author} {\bibfnamefont {Z.}~\bibnamefont {Drogosz}}, \bibinfo {author} {\bibfnamefont {W.}~\bibnamefont {Florkowski}}, \bibinfo {author} {\bibfnamefont {S.~K.}\ \bibnamefont {Kar}}, \ and\ \bibinfo {author} {\bibfnamefont {V.}~\bibnamefont {Mykhaylova}},\ }\href@noop {} {\  (\bibinfo {year} {2025})},\ \Eprint {http://arxiv.org/abs/2505.02657} {arXiv:2505.02657 [hep-ph]} \BibitemShut {NoStop}%
\bibitem [{\citenamefont {Mathisson}(1937)}]{Mathisson:1937zz}%
  \BibitemOpen
  \bibfield  {author} {\bibinfo {author} {\bibfnamefont {M.}~\bibnamefont {Mathisson}},\ }\href {https://www.actaphys.uj.edu.pl/fulltext?series=T&vol=6&no=3&page=163} {\bibfield  {journal} {\bibinfo  {journal} {Acta Phys. Polon.}\ }\textbf {\bibinfo {volume} {6}},\ \bibinfo {pages} {163} (\bibinfo {year} {1937})}\BibitemShut {NoStop}%
\bibitem [{\citenamefont {{Mathisson}}(2010)}]{2010GReGr..42.1011M}%
  \BibitemOpen
  \bibfield  {author} {\bibinfo {author} {\bibfnamefont {M.}~\bibnamefont {{Mathisson}}},\ }\href {\doibase 10.1007/s10714-010-0939-y} {\bibfield  {journal} {\bibinfo  {journal} {General Relativity and Gravitation}\ }\textbf {\bibinfo {volume} {42}},\ \bibinfo {pages} {1011} (\bibinfo {year} {2010})}\BibitemShut {NoStop}%
\bibitem [{\citenamefont {Jackson}(1998)}]{Jackson:1998nia}%
  \BibitemOpen
  \bibfield  {author} {\bibinfo {author} {\bibfnamefont {J.~D.}\ \bibnamefont {Jackson}},\ }\href@noop {} {\emph {\bibinfo {title} {Classical Electrodynamics}}}\ (\bibinfo  {publisher} {Wiley},\ \bibinfo {year} {1998})\BibitemShut {NoStop}%
\bibitem [{\citenamefont {Rajeev}()}]{Rajeev:2025QM}%
  \BibitemOpen
  \bibfield  {author} {\bibinfo {author} {\bibfnamefont {S.}~\bibnamefont {Rajeev}},\ }\href@noop {} {}\bibinfo {howpublished} {private communication, see also \url{https://indico.cern.ch/event/1334113/contributions/6291310/attachments/3044936/5379962/QM2025_Abboud_Poster.pdf}}\BibitemShut {NoStop}%
\bibitem [{\citenamefont {Frenkel}(1926)}]{Frenkel:1926zz}%
  \BibitemOpen
  \bibfield  {author} {\bibinfo {author} {\bibfnamefont {J.}~\bibnamefont {Frenkel}},\ }\href {\doibase 10.1007/BF01397099} {\bibfield  {journal} {\bibinfo  {journal} {Z. Phys.}\ }\textbf {\bibinfo {volume} {37}},\ \bibinfo {pages} {243} (\bibinfo {year} {1926})}\BibitemShut {NoStop}%
\bibitem [{\citenamefont {Drogosz}(2025)}]{Drogosz:2025ose}%
  \BibitemOpen
  \bibfield  {author} {\bibinfo {author} {\bibfnamefont {Z.}~\bibnamefont {Drogosz}},\ }\href@noop {} {\  (\bibinfo {year} {2025})},\ \Eprint {http://arxiv.org/abs/2504.03396} {arXiv:2504.03396 [hep-ph]} \BibitemShut {NoStop}%
\end{thebibliography}
\end{document}